\documentclass[amsmath,amssymb,prc,final,showpacs,twocolumn]{revtex4-1}
 
\usepackage{bm,hyphenat,xspace}
\usepackage{graphicx,epsfig}

\newcommand {\mbf}[1]{{\mathbf{#1}}}
\newcommand {\hv}[1]{{\hat{\mathbf{#1}}}}

\newcommand {\mcu}{\mathcal{U}}
\newcommand {\baru}{\bar{U}_1}
\newcommand {\mct}{\mathcal{T}}

\newcommand{\cm}{\mathrm{c\!\:\!.m\!\:\!.}}

\newcommand{\dd}{d\text{-}d}

\begin{document}

\title {Three-cluster breakup in deuteron-deuteron collisions: 
single-scattering approximation}
 
\author{A.~Deltuva} 
\affiliation
{Institute of Theoretical Physics and Astronomy, 
Vilnius University, A. Go\v{s}tauto St. 12, LT-01108 Vilnius, Lithuania
}

\author{A.~C.~Fonseca} 
\affiliation{Centro de F\'{\i}sica Nuclear da Universidade de Lisboa, 
P-1649-003 Lisboa, Portugal }

\received{\today}

\begin{abstract}
 We present  results for the three-cluster breakup in deuteron-deuteron
collisions at 130  and 270 MeV deuteron beam energy. 
The breakup amplitude is calculated using the
first term in the Neumann series expansion of the corresponding exact
four-nucleon equations. In analogy with nucleon-deuteron breakup
where an equivalent approximation is compared with exact calculations,
we expect this single-scattering approximation to provide a rough 
estimation of three-body  breakup observables in quasifree configurations. 
We predict the  nucleon-deuteron and deuteron-deuteron three-cluster breakup
cross sections to be of a comparable size and thereby question the reliability
of the recent experimental data 
[A. Ramazani-Moghaddam-Arani, Ph.D. thesis, University of Groningen, 2009;
A. Ramazani-Moghaddam-Arani {\it et~al.}, EPJ Web of Conferences {\bf 3},  04012  (2010)] 
that is smaller by about three orders of magnitude.
 We also show that an equivalent single-scattering approximation  
provides a reasonable description of deuteron-deuteron elastic scattering
at forward scattering angles.

\end{abstract}

 \maketitle

\section{Introduction \label{sec:intro}}

In the last 20 years experimentalists have provided precise data
for deuteron (d) breakup in the collision with a proton (p) 
at energies up to pion production
threshold. Their motivation has been to look for regions of three-body
phase space where the results are sensitive to the nucleon-nucleon
(NN) interaction and the need to include a three-nucleon (3N)
force. Given that there are numerically exact 3N
calculations that include a wide variety of NN force models,
comparison between theoretical calculations and precise data sheds
light on the quality of the NN interactions that have been
proposed. Since all NN interactions are fitted to NN elastic
scattering data with equivalent $\chi^2$ precision, possible disagreements
may reside in the off-shell nature of the chosen NN interaction or the
need to add a 3N force. Unfortunately this quest has been, in a few
isolated cases such as the nucleon-deuteron analyzing power puzzle
\cite{kievsky:01a} or the space-star anomaly
\cite{strate:89,setze:05a,deltuva:05d},
overshadowed by persistent discrepancies between data
and theory that defy comprehension within the framework of
non-relativistic quantum mechanics applied to three interacting
nucleons alone. 

Up to now very few breakup data exist involving four-nucleon
reactions, 
though this system shows extra sensitivity to NN force
models as some calculations have already demonstrated 
\cite{deltuva:07c,lazauskas:09a,viviani:13a}  for specific
spin observables and cross sections for elastic, charge exchange and
transfer reactions. Nevertheless  exact numerical four-nucleon
calculations for breakup amplitudes are still years away given the
dimensionality of the problem and the complex structure of
singularities above breakup. For elastic and transfer reactions driven
by all possible N+3N initial states as well as d+d, 
 converged solutions for all two-body reactions up to 30 MeV beam
energy \cite{deltuva:14a,deltuva:15a,lazauskas:15a} are available.
 Going to higher energies is formally not an obstacle but
requires much more computer power to obtain fully converged results
in terms of number of partial waves and grid points.  On the other
hand, the existing d+d breakup measurement
 \cite{ramazani:phd,dd3bkvi,PhysRevC.83.024002} 
performed at 130 MeV deuteron beam energy has provided fully
exclusive cross sections that are about three orders of magnitude smaller
than for the p+d breakup at comparable energies. This is
puzzling because one would expect a similar order of magnitude
for p+d and d+d  total breakup cross sections.
According to recent calculations, albeit at the much lower energy of 10 MeV
 \cite{deltuva:15a}, this is indeed the case.

Therefore, in order to provide the first estimate of d+d breakup cross
sections at higher energies, we present here an approximate calculation based on 
the lowest-order term
in the Neumann series expansion of the Alt, Grassberger and Sandhas
(AGS) equations \cite{grassberger:67}
 for the breakup operator and called single-scattering approximation (SSA).
 The SSA contains the breakup of one of the deuterons followed by the
scattering of one nucleon from the other deuteron through the full 3N
operator that sums up all orders of NN rescattering involving the
corresponding three particles. This operator is obtained from the exact
solution of the AGS three-nucleon equations for the underlying
force model being used. As demonstrated before \cite{dd232}, the same kind of approximation
 may be used to calculate d+d elastic scattering at high energies. 
In order to obtain a quantitative calibration of the d+d breakup cross sections, 
we provide SSA results also for d+d elastic scattering observables.

SSA for three-cluster breakup calculations is expected
to be reasonable near quasi-free
scattering (QFS) kinematics at energies  above 100 MeV in the
center-of-mass (c.m.) system. This conjecture is supported
 by the results for the equivalent
approximation applied to p+d breakup where a comparison with the exact
results that include NN rescattering to all orders is possible.

In Section II we present the theory for the breakup amplitudes using
the first term in the Neumann series expansion of the AGS equations,
and study the equivalent approximation applied to
p+d breakup by comparing such results with those obtained from the 
exact solution of 
the corresponding 3N problem. In Section III we show the results for
d+d elastic scattering and three-cluster breakup near  the quasi-free scattering kinematics.
 Conclusions come in Section IV.

\section{Scattering amplitudes \label{sec:eq}}

We start with exact four-nucleon scattering equations  and 
transition amplitudes, and then show how they are simplified 
under the single scattering approximation.
Treating nucleons as identical fermions in the isospin formalism,
one has to consider only two two-cluster partitions, namely,
 $3+1$ and $2+2$ which in the notation for operators and wave-functions
  are abbreviated by 1 and 2, respectively.
In terms of four particles 1,2,3, and 4 they correspond to the clustering
(12,3)4 and (12)(34), respectively. The reactions initiated by
the collisions of two deuterons
are described by the four-nucleon transition operators 
 $\mcu_{\beta 2}$ obeying the symmetrized AGS equations 
\begin{subequations}  \label{eq:AGS}   
\begin{align}  
\mcu_{12}  = {}&  (G_0 \, t \, G_0)^{-1} -
P_{34} U_1 G_0 \, t \, G_0 \, \mcu_{12}  
+ U_2   G_0 \, t \, G_0 \, \mcu_{22}, \label{eq:U11}  \\
\label{eq:U21}
\mcu_{22} = {}&  (1 - P_{34}) U_1 G_0 \, t \, G_0 \, \mcu_{12}.
\end{align}
\end{subequations}
The pair potential $v$ enters the AGS equations via the
pair (12) transition matrix
\begin{gather} \label{eq:t}
t = v + v G_0 t
\end{gather} 
and via the  3+1 and 2+2 subsystem transition operators 
\begin{gather} \label{eq:AGSsub}
U_\alpha =  P_\alpha G_0^{-1} + P_\alpha t\, G_0 \, U_\alpha.
\end{gather}
The dependence of all transition operators on the available energy $E$, 
although not indicated in our notation, arises via the free resolvent 
\begin{gather}\label{eq:G0}
G_0 = (E +i0 - H_0)^{-1}
\end{gather} 
where  $H_0$ is the  free Hamiltonian.
In the considered case $E = p_2^2/2m_N + 2 \epsilon_d$, where
$p_2$ is the relative $\dd$ momentum, $m_N$ the average nucleon mass,
 and $\epsilon_d = -2.225$ MeV the energy of the deuteron bound state.
The full antisymmetry that a system of four identical fermions must obey is ensured by the 
permutation operators $P_{ab}$ of nucleons $a$ and $b$ with
$P_1 =  P_{12}\, P_{23} + P_{13}\, P_{23}$ and $P_2 =  P_{13}\, P_{24}$.
Furthermore,
the basis states are antisymmetric under exchange of the two nucleons 
(12) while in the $2+2$ partition they are antisymmetric
also  under exchange of the two nucleons (34).

The three-cluster breakup operator for the $2+2$ collision is derived in 
Ref.~\cite{deltuva:12e} to be
\begin{equation} \label{eq:U3}
\mcu_{32} = (1- P_{34}) U_1 G_0 \, t \, G_0 \, \mcu_{12} +  
U_2 G_0 \,  t \, G_0 \, \mcu_{22} .
\end{equation}

To get the reaction amplitudes the on-shell matrix elements of
the above transition operators have to be calculated between the appropriate 
channel states. Considering  
only the elastic $\dd$ scattering and three-cluster breakup,
the respective channel states are
 $| \Phi_{2}(\mbf{p}_2) \rangle = | \phi_{d}  \phi_{d}  \mbf{p}_2 \rangle = 
(1+P_2)| \phi_{2}(\mbf{p}_2) \rangle$
with the Faddeev component
$|\phi_{2}(\mbf{p}_2) \rangle = G_0 v | \Phi_{2}(\mbf{p}_2) \rangle =
G_0 t P_2 |\phi_{2} (\mbf{p}_2) \rangle$,
and  $| \Phi_{3}(\mbf{k}_y,\mbf{k}_z) \rangle = 
| \phi_{d} \mbf{k}_y  \mbf{k}_z \rangle$.
Here $ \phi_{d}$ denotes the deuteron wave function,
$\mbf{p}_2$ is the relative $\dd$ momentum, and
$\mbf{k}_y$ and $\mbf{k}_z$ are  momenta for the relative motion 
in the three-cluster $d+p+n$ system, e.g.,
$\mbf{k}_y = (2\mbf{k}_p-\mbf{k}_d)/3$ and 
$\mbf{k}_z= (3\mbf{k}_n-\mbf{k}_p-\mbf{k}_d)/4$ 
for the $(dp)n$ partition. 
 The dependence on the
discrete spin and isospin quantum numbers is suppressed in our notation.
The elastic and three-cluster breakup amplitudes are
\begin{subequations}  \label{eq:ampl}   
\begin{align}  
\langle \mbf{p}'_2 | \mct_{22} | \mbf{p}_2 \rangle
= {}& 2 \langle \phi_{2}(\mbf{p}'_2)| \mcu_{22} | \phi_{2}(\mbf{p}_2) \rangle, \\
\langle \mbf{k}_y\mbf{k}_z |\mct_{32} | \mbf{p}_2 \rangle
= {}& 2 \langle \Phi_{3}(\mbf{k}_y,\mbf{k}_z)|\mcu_{32} 
| \phi_{2}(\mbf{p}_2) \rangle.
\end{align}
\end{subequations}
Factors of 2 result from the antisymmetrization 
of the four-nucleon states \cite{deltuva:12e}; to account for the
identity of the two deuterons one has to use  symmetrized
$\dd$ channel states, leading to symmetrized amplitudes
obtained from \eqref{eq:ampl}   by replacing
$| \phi_{2}(\mbf{p}_2) \rangle$ with 
$ | \phi_{2}^s(\mbf{p}_2) \rangle =
 \frac{1}{\sqrt{2}}(1+P_2) | \phi_{2}(\mbf{p}_2) \rangle
= \frac{1}{\sqrt{2}} [| \phi_{2}(\mbf{p}_2) \rangle + | \phi_{2}^x(-\mbf{p}_2) \rangle] $
where the superscript $x$ indicates that also the spin-isospin part 
is exchanged. Note that in the partial-wave representation
both terms have equal contributions.

\begin{figure}[!]
\begin{center}
\includegraphics[scale=0.62]{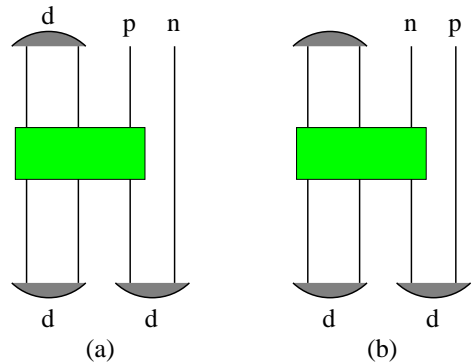}
\end{center} 
\caption{ \label{fig:u} (Color online) 
Two contributions to the single-scattering three-cluster
breakup amplitude. The three-nucleon transition operator $U_1$
is represented by a box while deuterons 
are represented by filled arcs. }
\end{figure}

The absence of an inhomogeneous term in the equation \eqref{eq:U21} makes possible the 
development of a simple approximation for d+d elastic scattering and 
three-cluster breakup that may 
work at higher energies under certain conditions. In the single-scattering 
approximation  only the terms that are of 
first order in the subsystem transition operators $U_\alpha$
are retained, resulting in the amplitudes
\begin{subequations}  \label{eq:ssa}   
\begin{align}   \label{eq:ssa-a}   
\langle \mbf{p}'_2 | \mct_{22}^{\rm SS} | \mbf{p}_2 \rangle
= {}& 2  \langle \phi_{2}^s(\mbf{p}'_2)| (1 - P_{34}) U_1
 | \phi_{2}^s(\mbf{p}_2) \rangle, \\ \label{eq:ssa-b}   
\langle \mbf{k}_y\mbf{k}_z | \mct_{32}^{\rm SS} | \mbf{p}_2 \rangle
= {}& 2 \langle \Phi_{3}(\mbf{k}_y,\mbf{k}_z)| (1 - P_{34}) U_1
| \phi_{2}^s(\mbf{p}_2) \rangle.
\end{align}
\end{subequations}
Thus, only the calculation of the operator $U_1$ is required,
which is the standard three-nucleon AGS transition matrix.
However, it needs to be calculated at off-shell momenta
as we show below.
Given that 
$|\phi_{2}(\mbf{p}_2) \rangle = G_0 v | \phi_{d}  \phi_{d}  \mbf{p}_2 \rangle $, 
it is convenient to introduce the momentum-space matrix elements
\begin{equation} \label{eq:U1yz}
\langle  \phi_{d} \mbf{k}_y\mbf{k}_z | v G_0 U_1 G_0 v | 
 \phi_{d} \mbf{k}'_y\mbf{k}'_z  \rangle =
\delta(\mbf{k}_z - \mbf{k}'_z) \baru(\mbf{k}_y,\mbf{k}'_y,{k}_z)
\end{equation}
and the deuteron wave function $\phi_{d}(\mbf{p}_y)= \langle \mbf{p}_y | \phi_{d} \rangle$; 
they are  operators in the spin-isospin space.
With these definitions the first contribution to the symmetrized SSA elastic
amplitude \eqref{eq:ssa} is calculated as
\begin{gather}\label{eq:sse}
\begin{split}
\langle \phi_{2}(\mbf{p}'_2){}& | (1 -  P_{34}) U_1
 | \phi_{2}(\mbf{p}_2) \rangle \\ &
=  2 \int d^3 \mbf{k}_z \phi_{d}(\mbf{p}'_y) \baru(\mbf{k}_y,\mbf{k}'_y,{k}_z) \phi_{d}(\mbf{p}_y)
\end{split}
\end{gather} 
with
\begin{subequations}  \label{eq:ssep}   
\begin{align}  
\mbf{p}'_y = {}& \frac12 \mbf{p}'_2 - \mbf{k}_z, \\
\mbf{p}_y = {}& \frac12 \mbf{p}_2 - \mbf{k}_z, \\
\mbf{k}'_y = {}& \mbf{p}_2 - \frac23 \mbf{k}_z, \\
\mbf{k}_y = {}& \mbf{p}'_2 - \frac23 \mbf{k}_z.
\end{align}
\end{subequations}
The remaining three contributions are calculated analogously.
In Eq.~\eqref{eq:sse} the energy available for the three-nucleon subsystem
$E- 2k_z^2/3m_N$ runs from $E$ to $-\infty$, obviously indicating
that $U_1$ is needed off-shell. On the other
hand, the final breakup channel state 
$| \Phi_{3}(\mbf{k}_y,\mbf{k}_z) \rangle$ 
fulfills the on-shell condition $E = \epsilon_d + 3k_y^2/4m_N + 2k_z^2/3m_N$,
indicating that only half-shell elements of  $U_1$ are needed.
Thus,  the first term of the single-scattering breakup
amplitude \eqref{eq:ssa-b}  becomes
\begin{gather}\label{eq:ssb}
\langle \Phi_{3}(\mbf{k}_y,\mbf{k}_z)|  U_1
| \phi_{2}(\mbf{p}_2) \rangle=
 \baru(\mbf{k}_y,\mbf{k}'_y,{k}_z) \phi_{d}(\mbf{p}_y)
\end{gather} 
with $\mbf{k}'_y$ and $\mbf{p}_y$ defined in Eqs.~\eqref{eq:ssep}.
The second term is calculated analogously, but with the final state
$P_{34}| \Phi_{3}(\mbf{k}_y,\mbf{k}_z) \rangle = 
|\Phi^x_{3}(\frac13 \mbf{k}_y+ \frac89 \mbf{k}_z,\mbf{k}_y - \frac13 \mbf{k}_z) \rangle  $
where the superscript $x$ indicates that also the spin-isospin part 
is exchanged.
These two contributions are graphically depicted in 
Fig.~\ref{fig:u} for the $d+p+n$ final state. 
Taking the left-side deuteron as the beam and the right-side deuteron
as the target, the diagram (a) corresponds to the target
deuteron breakup after the
full interaction between the impinging deuteron and the target proton 
while no interaction occurs involving the target neutron.
Thus, the diagram (a) corresponds to proton-deuteron
quasi-free scattering (QFS). 
Analogously, the diagram (b) corresponds to neutron-deuteron QFS.
Two more contributions 
$\langle \Phi_{3}(\mbf{k}_y,\mbf{k}_z)| (1-P_{34}) U_1| \phi_{2}^x(-\mbf{p}_2) \rangle$,
not shown in Fig.~\ref{fig:u},
 arise due to the symmetrization of the initial
$d+d$ state; they correspond to the breakup of the impinging deuteron.

Under the assumption of the simplified SSA reaction
mechanism of Fig.~\ref{fig:u} (a),
the energy distribution of the final neutron is given by the 
deuteron wave function, i.e., the differential cross section is sharply peaked at the neutron
energy $E_n = 0$. In a complete reaction picture the  cross section
 also gets contributions from the higher rescattering terms beyond the SSA; roughly,
their relative importance increases when the SSA contribution decreases, i.e.,
for larger $E_n$. Thus, the reliability of the SSA should decrease with
increasing energy of the final neutron. 
 Another necessary condition for the non-interacting neutron, and thereby
also for the validity of the SSA, 
is a high enough relative $n$-$d$ and $n$-$p$ energy, implying
also high enough energy for the initial beam and for the final deuteron and proton.
At  $E_n = 0$ only the contribution of Fig.~\ref{fig:u} (a)
is peaked; the remaining three SSA contributions are not as they 
do not correspond to the deuteron-proton(target) QFS. 
In the following we will call the results obtained with only
the contribution of Fig.~\ref{fig:u} (a) as SSA-1, while those
including all four contributions as SSA-4. An agreement between the
SSA-1 and  SSA-4 results indicates the dominance of the Fig.~\ref{fig:u} (a)
SSA reaction mechanism,  while  disagreement indicates a more complicate reaction 
mechanism; a significant contribution of higher-order terms
is probable in the latter case although it cannot be ruled out also in the former case. 

\begin{figure}[!]
\begin{center}
\includegraphics[scale=0.62]{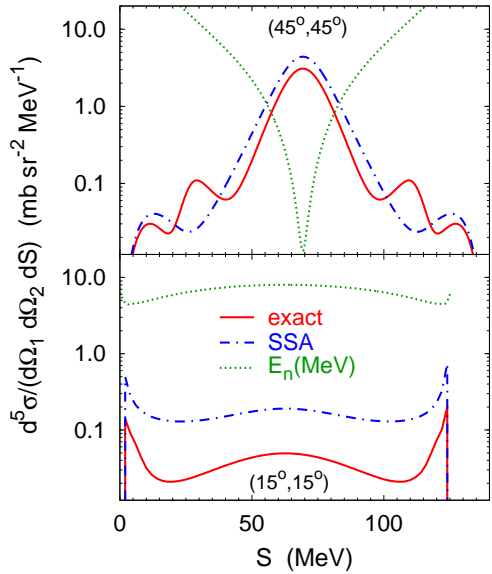}
\end{center} 
\caption{ \label{fig:ss3-95} (Color online) 
Differential cross section for the proton deuteron breakup
at 95  MeV proton beam energy as function of the arclength $S$
along the kinematical curve. The polar angles of the  two detected protons 
$(\theta_1,\theta_2)$ are indicated in the plots.
Exact results based on the CD Bonn + $\Delta$ potential
and including the Coulomb interaction (solid curves) are compared
with SSA results (dashed-dotted curves) that neglect the Coulomb interaction.
The final neutron energy is given by dotted curves. }
\end{figure}

\begin{figure}[!]
\begin{center}
\includegraphics[scale=0.62]{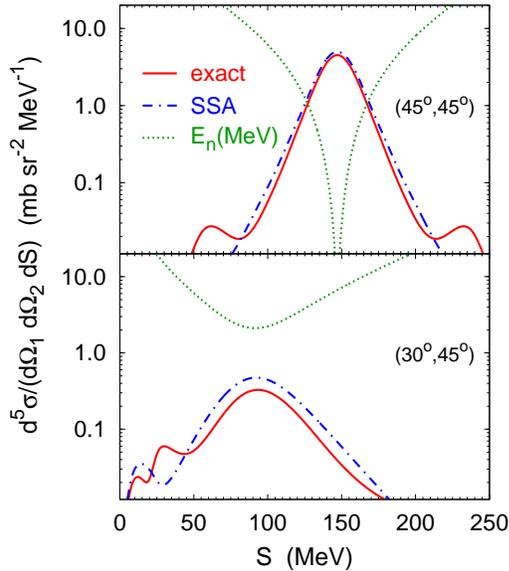}
\end{center} 
\caption{ \label{fig:ss3-200} (Color online) 
Same as Fig.~\ref{fig:ss3-95} but
at 200  MeV proton beam energy.}
\end{figure}

Finally we note that the corresponding SSA can be introduced also in the 
nucleon-deuteron scattering, expanding the three-nucleon transition operators in terms
of two-nucleon transition operators $t$ and retaining first order terms in the Neumann series.
 In fact, this approach already has been used in a number 
of early works, e.g. \cite{fukehara:79}.
The full nucleon-deuteron  breakup operator is 
\begin{gather} \label{eq:U0}
U_0 =  (1+P_1) t\, G_0 \, U_1
\end{gather}
whereas its SSA reads
\begin{gather} \label{eq:U0ssa}
U_0^{\rm SS} =  (1+P_1) t\, P_1.
\end{gather}
The diagrammatic representation of the nucleon-deuteron SSA is
very similar to the one of Fig.~\ref{fig:u} except that the 
left-side deuteron is replaced by a nucleon and the three-nucleon
transition operator $U_1$ is replaced by the two-nucleon operator $t$.

Comparing results based on Eqs.~\eqref{eq:U0} and \eqref{eq:U0ssa}
one can evaluate the reliability of the SSA in three-nucleon breakup.
This is done in Figs.~\ref{fig:ss3-95} and \ref{fig:ss3-200}
for proton-deuteron breakup at 
95 and 200 MeV proton beam energy in several kinematical configurations.
The above energy values approximately correspond to the same c.m. energy as 
130 and 270 MeV deuteron  energy in deuteron-deuteron collisions.
The fivefold differential cross section
$d^5\sigma/d\Omega_1 d\Omega_2 dS$ is shown as a function of the arclength $S$
along the kinematical curve
for fixed angles of the two detected protons $(\theta_1,\varphi_1=0^\circ)$
and $(\theta_2,\varphi_2=180^\circ)$. The energy of the neutron $E_n$ is plotted 
as well. One can see that the SSA results deviate significantly
from the exact ones for larger $E_n$ values; for small $E_n$ well below 1 MeV 
the agreement improves for higher  proton beam energy.

\begin{figure}[!]
\begin{center}
\includegraphics[scale=0.56]{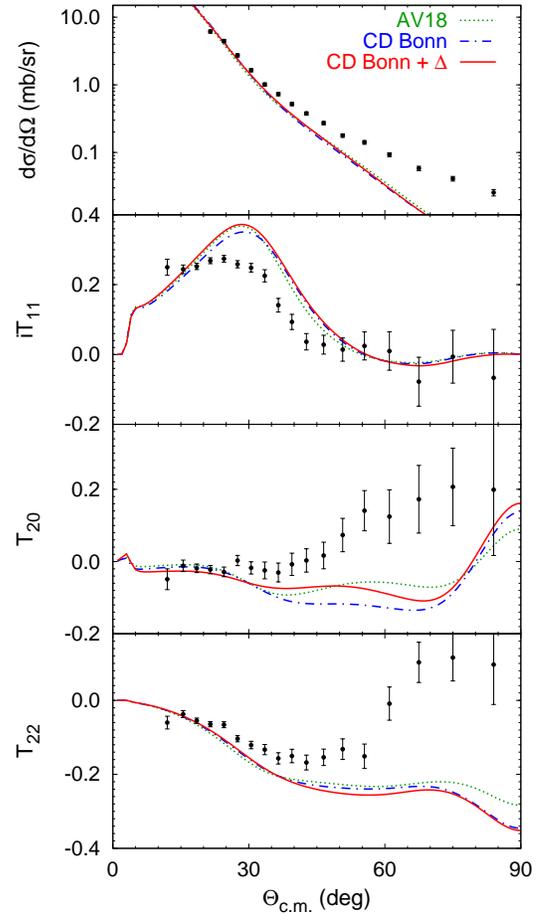}
\end{center} 
\caption{ \label{fig:dd232} (Color online) 
Differential cross section and deuteron analyzing powers
for the deuteron-deuteron elastic scattering
at 232  MeV deuteron beam energy.  
SSA predictions based on the CD Bonn + $\Delta$ (solid curves),
CD Bonn (dashed-dotted curves), and AV18  (dotted curves)
potentials are compared with experimental data from
Ref.~\cite{dd232}. }
\end{figure}

Total p+d breakup cross sections calculated exactly and in the SSA are
69.1 and 109 mb at 95 MeV and 49.7 and 63.5 mb at 200 MeV, respectively. 
Thus, SSA provides a correct order of the magnitude
for the p+d total  breakup cross section.

\section{Results \label{sec:res}}

\begin{figure*}[!]
\begin{center}
\includegraphics[scale=0.7]{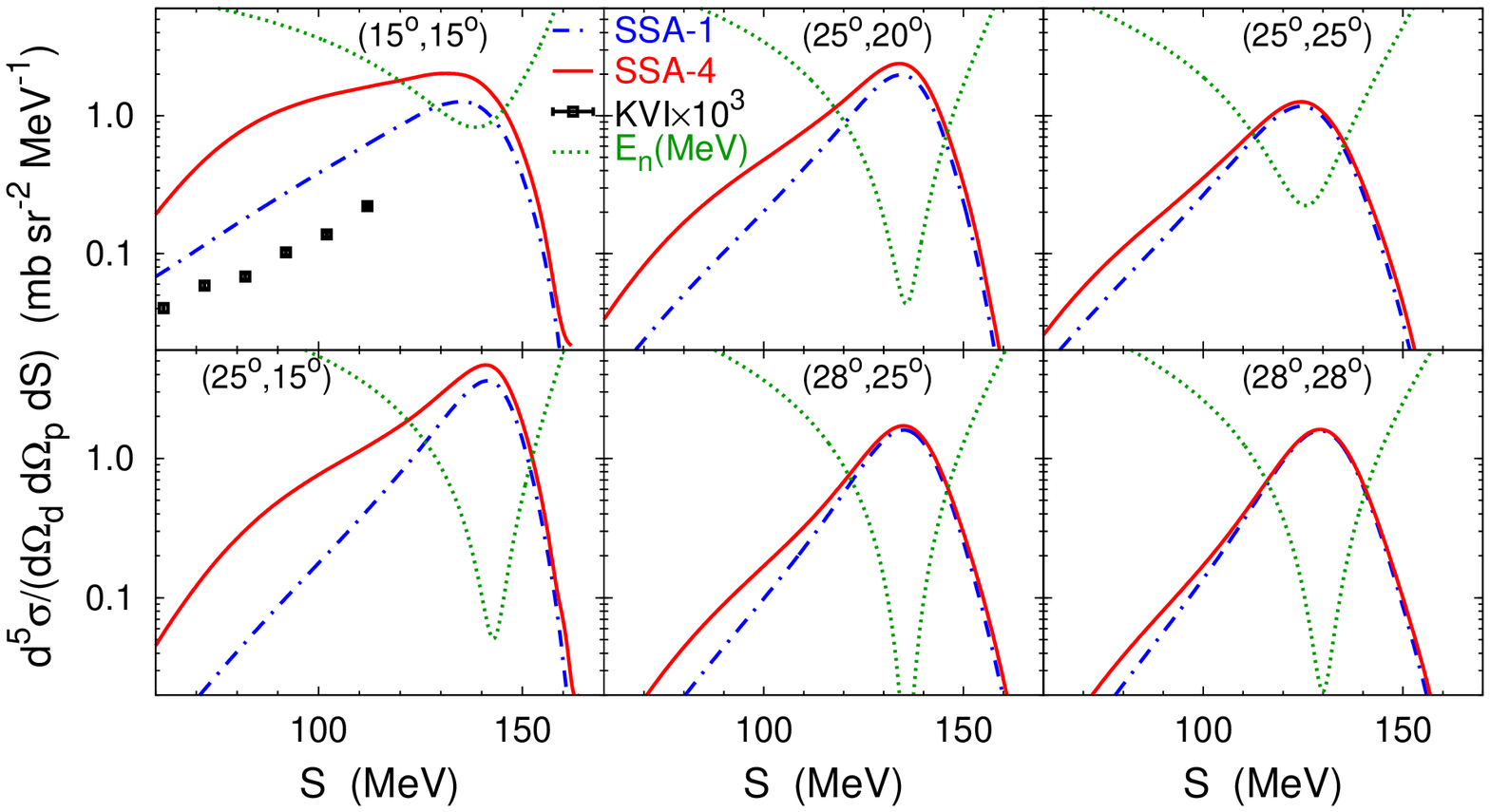}
\end{center} 
\caption{ \label{fig:dd130} (Color online) 
Differential cross section for the three-cluster breakup
in deuteron-deuteron collision
at 130  MeV deuteron beam energy as function of the arclength $S$
along the kinematical curve. The polar angles of the detected deuteron and proton
$(\theta_d,\theta_p)$ are indicated in the plots, the azimuthal angles 
are $(\varphi_d,\varphi_p) = (0^\circ,180^\circ)$.
SSA 4-term (solid curves) and 1-term (dashed-dotted curves)
results, both based on the CD Bonn + $\Delta$ potential, are compared.
Published experimental data from Ref.~\cite{dd3bkvi} 
are only available for the configuration $(15^\circ,15^\circ)$;
in the plot the data points are multiplied by a factor of $10^3$.
The final neutron energy is given by  dotted curves. }
\end{figure*}

\begin{figure}[!]
\begin{center}
\includegraphics[scale=0.66]{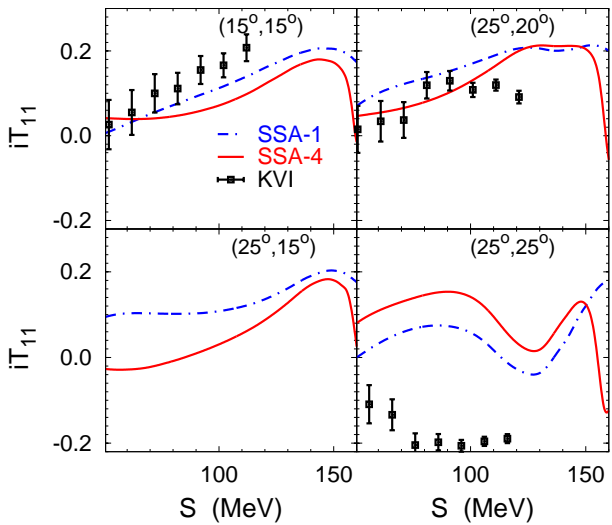}
\end{center} 
\caption{ \label{fig:t11} (Color online) 
Deuteron analyzing power $iT_{11}$ for the three-cluster breakup
in deuteron-deuteron collision at 130  MeV deuteron beam energy. 
 The polar angles of the detected deuteron and proton
$(\theta_d,\theta_p)$ are indicated in the plots.
SSA 4-term (solid curves) and 1-term (dashed-dotted curves)
results, both based on the CD Bonn + $\Delta$ potential, are compared
with the experimental data from Refs.~\cite{PhysRevC.83.024002,ramazani:phd}.
 }
\end{figure}

\begin{figure}[!]
\begin{center}
\includegraphics[scale=0.66]{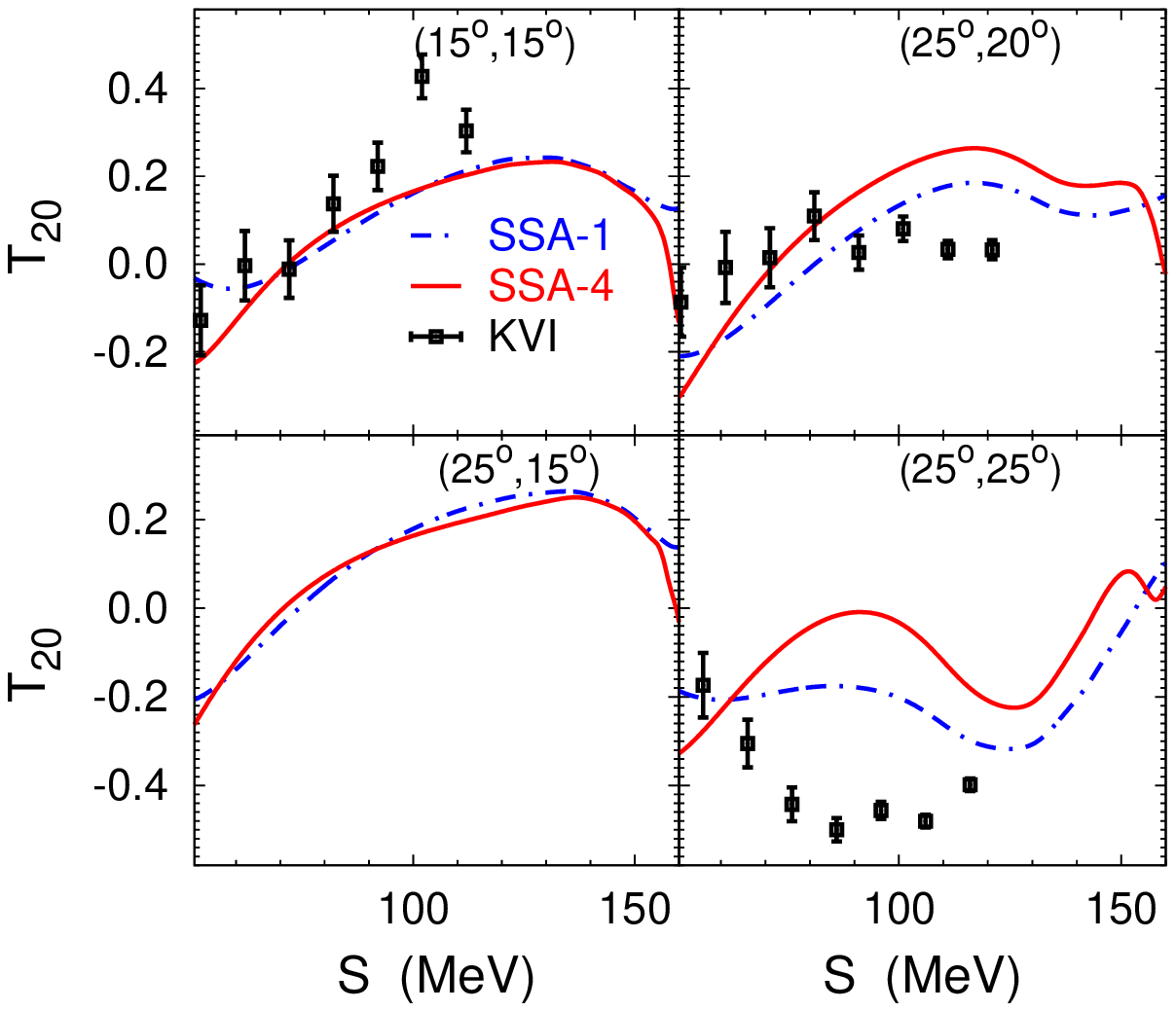}
\end{center} 
\caption{ \label{fig:t20} (Color online) 
Deuteron tensor analyzing power $T_{20}$ for the three-cluster breakup
in deuteron-deuteron collision at 130  MeV deuteron beam energy. 
Curves and experimental data are as in Fig.~\ref{fig:t11}.}
\end{figure}

\begin{figure}[!]
\begin{center}
\includegraphics[scale=0.66]{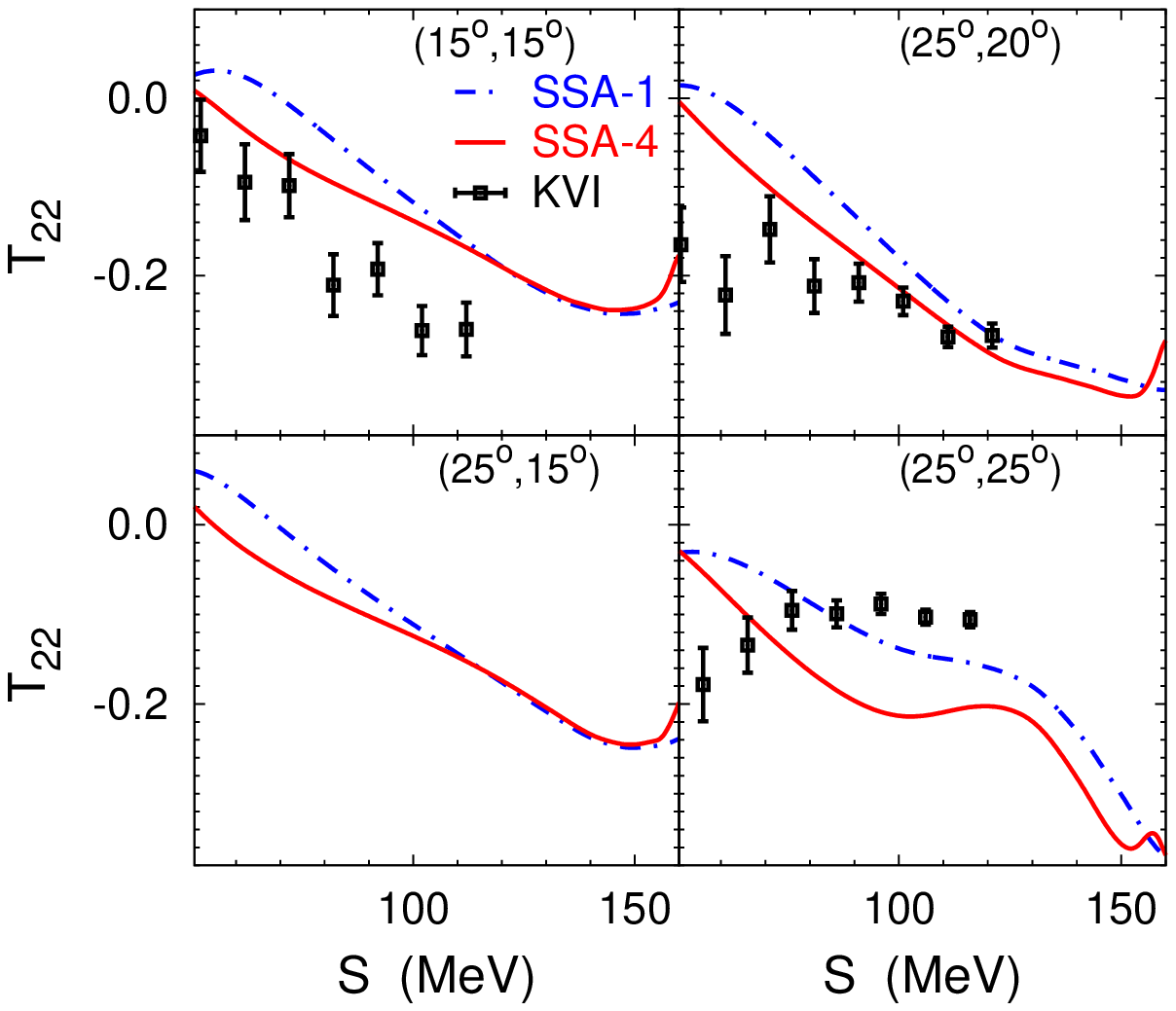}
\end{center} 
\caption{ \label{fig:t22} (Color online) 
Deuteron tensor analyzing power $T_{22}$ for the three-cluster breakup
in deuteron-deuteron collision at 130  MeV deuteron beam energy. 
Curves and experimental data are as in Fig.~\ref{fig:t11}.}
\end{figure}

\begin{figure}[!]
\begin{center}
\includegraphics[scale=0.6]{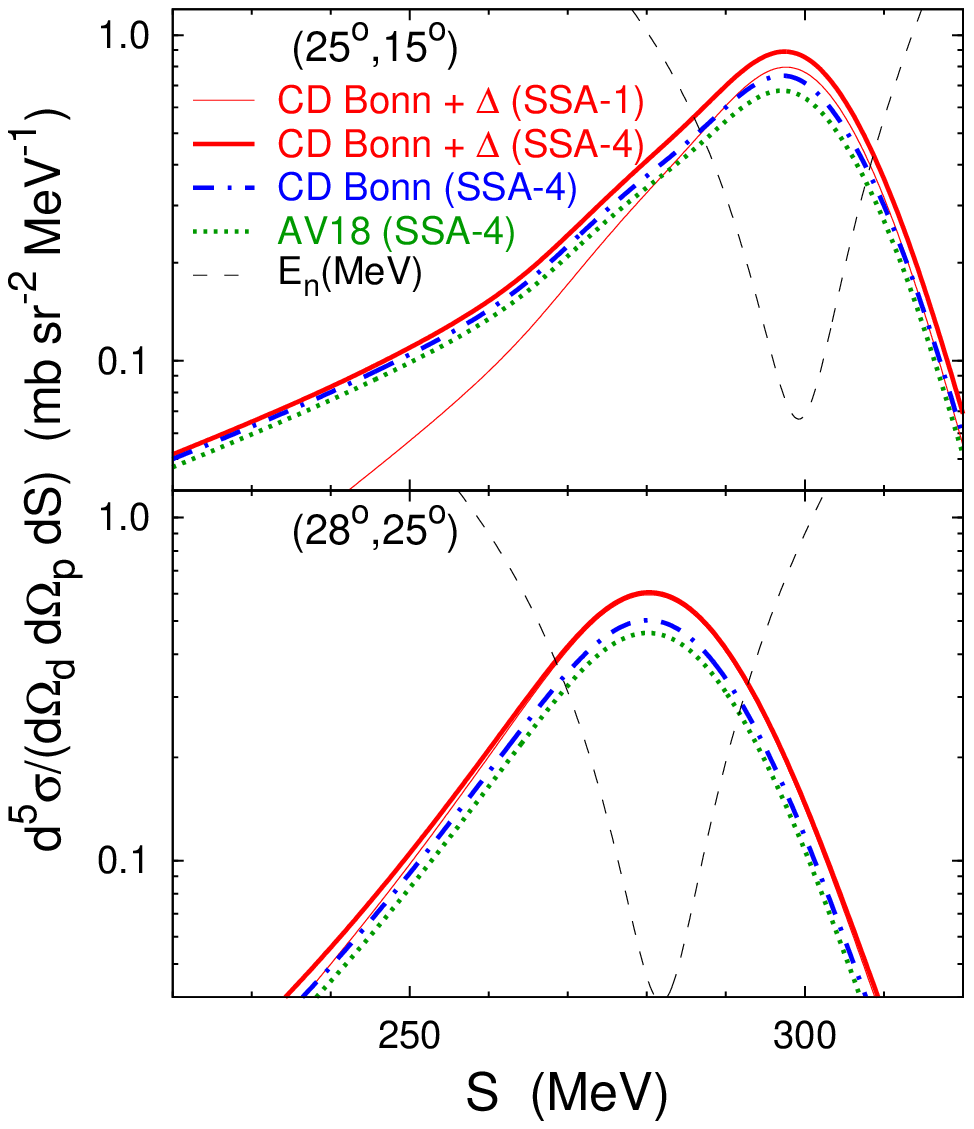}
\end{center} 
\caption{ \label{fig:dd270v} (Color online) 
Differential cross section for the three-cluster breakup
in deuteron-deuteron collision
at 270  MeV deuteron beam energy.  The polar angles of the detected deuteron and proton
$(\theta_d,\theta_p)$ are indicated in the plots.
SSA-4 based on the CD Bonn + $\Delta$ (solid curves),
CD Bonn (dashed-dotted curves), and AV18  (dotted curves)
potentials are compared with the  SSA-1 predictions 
 based on the CD Bonn + $\Delta$ potential (thin solid curves).
The final neutron energy is given by dashed curves. }
\end{figure}

We use three models of  realistic high-precision
two-nucleon potentials, namely,
the Argonne 18-operator (AV18) potential  \cite{wiringa:95a},
the charge-dependent Bonn potential (CD Bonn)  \cite{machleidt:01a},
and its extension CD Bonn + $\Delta$ \cite{deltuva:03c}
allowing for an excitation of a nucleon to a $\Delta$ isobar
and thereby yielding an effective 3N force. 

The three-nucleon AGS equation \eqref{eq:AGSsub} is solved in the
momentum-space partial-wave representation.
The states with two-nucleon angular momenta $j_x  \le 5$ and
 three-nucleon angular momenta $j_y  \le \frac{27}{2}$ are included.
The final four-nucleon results are well converged with these cutoffs.
The elastic amplitude \eqref{eq:ssa-a} is calculated in partial waves 
including the states with 4N total angular momentum $J \le 25$ while  
for the breakup amplitude 
the obtained partial-wave half-shell matrix elements of $U_1$ are first transformed into
the plane-wave basis and then used in Eq.~\eqref{eq:ssa-b}.
The Coulomb interaction is neglected
when solving AGS equation \eqref{eq:AGSsub} for the 3N transition operator $U_1$.
However, the external Coulomb correction is added to the 
elastic amplitude \eqref{eq:ssa-a}.

We first present results for d+d elastic observables at 232 MeV
beam energy where the experimental data \cite{dd232} for the differential cross section
and analyzing powers are available. The SSA results based on Eq.~\eqref{eq:ssa-a}   
 are shown in Fig.~\ref{fig:dd232}  for CD Bonn + $\Delta$, CD Bonn and AV18
potentials. The sensitivity to the force model is visible for the tensor analyzing powers
$T_{20}$ and $T_{22}$. The observables are reasonably described at forward scattering angles 
$\Theta_{\cm} \le 40^\circ$ in the c.m. frame. 
At larger scattering angles multiple scattering effects are expected, thus,
the failure of SSA is not surprising.
For the vector analyzing
 power $iT_{11}$ a rough agreement between SSA results and experimental data
is seen also at larger angles. Since the
calculation of the d+d elastic amplitude involves the integration of
the N+d off-shell amplitude over the 3N subsystem energy range from $E$ to $-\infty$, it is
reasonable to assume that the SSA for the d+d breakup may have the correct
order of magnitude over a wide range of phase space. Therefore we
expect our SSA results to provide at least the
correct order of magnitude for the d+d three-cluster breakup cross sections.

The fivefold differential cross section for the three-cluster breakup in the laboratory frame
\begin{gather}\label{eq:d5S} 
\begin{split}
   \frac{d^5 \sigma}{d\Omega_d d\Omega_p dS} = {}& (2 \pi)^{4}  \frac{ m_N }{p_2}
   \big | \langle \mbf{k}_y\mbf{k}_z | \mct_{32}^{\rm SS} | \mbf{p}_2 \rangle \big |^2  \, 
   m_N^2  k_d^2  k_p^2   \\ & \times
    \Big \{ (k_d/2)^2 \big[ 2k_p  - \hv{k}_p \cdot (2\mbf{p}_2  \!-\! \mbf{k}_d) \big]^2 \\ & +
k_p^2 \big[ 3k_d/2  - \hv{k}_d \cdot (2\mbf{p}_2  \!-\! \mbf{k}_p) \big]^2   \Big \}^{-1/2}.
  \end{split}
\end{gather}
is calculated as a function of the arclength $S$
along the kinematical curve in the plane of final deuteron and proton energies
$E_d$ and $E_p$ 
for fixed polar and azimuthal angles of the deuteron  $(\theta_d,\varphi_d=0^\circ)$
and proton $(\theta_p,\varphi_d=180^\circ)$. The starting point $S=0$ 
 is chosen as  $E_p = 0$ with $dE_p/dE_d > 0$ and $S$ is measured counterclockwise
in the $(E_d,E_p)$ plane.

The spin-averaged differential cross section for
six angular configurations is shown in Fig.~\ref{fig:dd130}, together with the
energy $E_n$ of the final-state neutron; the CD Bonn + $\Delta$ model is used.
This observable in peaked at the minimum of $E_n$,
i.e., near/at deuteron-proton QFS kinematics. The SSA-1 and SSA-4 predictions 
approach each other for small  $E_n$ and deviate at larger $E_n$.
The experimental data \cite{ramazani:phd}  taken at KVI are preliminary and cannot be shown
here except for the configuration  $(\theta_d,\theta_p) = (15^\circ,15^\circ)$ already
published in Ref.~\cite{dd3bkvi}. The disagreement between our predictions and data 
exceeds a factor of 1000 which is striking. In fact,  when compared to the 
data \cite{ramazani:phd}, a  similar discrepancy of three orders
of magnitude  exists for all remaining configurations in Fig.~\ref{fig:dd130}. 
Such a difference is far too large even taking into account that 
our predictions are approximate. On the other hand, using SSA we predict the same order of magnitude
for the differential cross section in nucleon-deuteron and deuteron-deuteron three-cluster breakup.
This applies also to the total three-cluster breakup cross section. Using
SSA-4 with the CD Bonn + $\Delta$ potential we obtain
 189 and 57 mb at 130 and 270 MeV, respectively, which are comparable to the ones 
in nucleon-deuteron breakup. Since in the nucleon-deuteron breakup the SSA provides the right order 
of magnitude for both differential and total cross sections, we believe that our
predictions for the deuteron-deuteron three-cluster breakup are reasonable as well and point out
to problems in the experimental data of Ref.~\cite{ramazani:phd}.
These conclusions are supported by very recent measurements of this reaction 
at 160 MeV deuteron beam energy where our differential cross section predictions 
near QFS kinematics are in qualitative 
agreement with the preliminary experimental data  \cite{khatri:phd}.

Deuteron vector  analyzing power $iT_{11}$ and 
tensor analyzing powers $T_{20}$ and $T_{22}$ are shown in Figs.~\ref{fig:t11} --- \ref{fig:t22}
for several kinematic $(\theta_d,\theta_p)$ configurations. The experimental data points for two of them, 
$(15^\circ,15^\circ)$ and $(25^\circ,25^\circ)$,
are taken from Ref.~\cite{PhysRevC.83.024002}, others are still preliminary and 
only available in Ref.~\cite{ramazani:phd}.  
There is a rough qualitative agreement between the data and SSA-1 and SSA-4 predictions
for $(15^\circ,15^\circ)$ and $(25^\circ,20^\circ)$ configurations.
The data change very rapidly from $(25^\circ,20^\circ)$ to $(25^\circ,25^\circ)$, loosing the 
agreement with SSA results. Comparing  SSA-1 and SSA-4 predictions one may conclude
that the level of agreement between SSA-1 and SSA-4 is not the same
 for the spin-averaged differential cross section and for spin observables. 
The sensitivity of all these observables to the force model is minor and it is not shown.

Although no experimental data is available above 160 MeV, we present in Fig.~\ref{fig:dd270v} example 
results at 270 MeV deuteron beam energy where the sensitivity to the force model becomes more visible. 
Under suitable kinematic conditions, e.g., in the $(28^\circ,25^\circ)$ configuration, 
this sensitivity even exceeds the  difference between SSA-1 and SSA-4.
Nevertheless it is plausible that this extra sensitiveness may change when exact calculations are performed.

\section{Summary \label{sec:sum}}

In the present work we have calculated the spin-averaged fivefold differential cross section
and deuteron analyzing powers for the three-cluster breakup in deuteron-deuteron
collisions at 130 and 270 MeV beam energy. 
Although at this time we
could not perform an exact four-nucleon calculation of the corresponding
breakup amplitudes, we devised an approximation
based on the first term in the Neumann series expansion of the AGS
three-cluster breakup operator which is expected to be, qualitatively, a reasonable
approximation near proton-deuteron quasifree scattering kinematic conditions.

In order to validate the method, we calculated proton-deuteron breakup at similar
energies and compared with the results of an exact three-nucleon
calculation. Likewise we used the same single scattering approximation
to calculate deuteron-deuteron elastic scattering at 232 MeV where there is data for
the differential cross section and analyzing powers. These auxiliary studies have shown that the
SSA  reproduces, if nothing else, the correct magnitude
of the existing cross section data.

Although the deuteron beam energy of 130 MeV seems to be not high enough for 
SSA to be reliable,  we compared
the SSA results for the three-cluster breakup
with the experimental data measured at KVI \cite{ramazani:phd,dd3bkvi,PhysRevC.83.024002}.
We found, at least in some configurations, a rough qualitative agreement
for deuteron analyzing powers, but a factor of 1000 difference for the cross section.
Since we have provided convincing arguments that SSA should yield at least correct order of
magnitude for total and differential cross sections, and that deuteron-deuteron and proton-deuteron
breakup cross sections should be of comparable size, our work raises some serious concern
on the correct normalization of the KVI data. On the contrary, new  Cracow
data \cite{khatri:phd} at 160 MeV seems to be in line with our SSA results.
Further investigation on these issues needs to be
pursued by both theory and experimental collaborations.

\vspace{3mm}

We thank A. Ramazani-Moghaddam-Arani and N. Kalantar-Nayestanaki for discussions and  providing
the experimental data.


\end{document}